\documentclass[12pt]{article}
\usepackage{epsf,here,amssymb}
\def\tfrac#1#2{{\textstyle{#1\over #2}}}
\def\half{{\tfrac{1}{2}}}
\def\barr{\begin{array}}
\def\earr{\end{array}}
\def\bfone{\relax{\rm 1\kern-.35em 1}}

\def\IE{\relax{{\rm I\kern-.18em E}}}
\def\IGam{\relax{{\rm I}\kern-.18em \Gamma}}

\def\IA{\relax{\hbox{{\rm A}\kern-.82em {\rm A}}}}

\def\IP{\relax{\rm I\kern-.18em P}}

\font\cmss=cmss10 \font\cmsss=cmss10 at 7pt

\def\inbar{\vrule height1.5ex width.4pt depth0pt}
\def\IC{\relax\,\hbox{$\inbar\kern-.3em{\rm C}$}}
\def\IG{\relax\,\hbox{$\inbar\kern-.3em{\rm G}$}}
\def\IB{\relax{\rm I\kern-.18em B}}
\def\ID{\relax{\rm I\kern-.18em D}}
\def\IL{\relax{\rm I\kern-.18em L}}
\def\IE{\relax{\rm I\kern-.18em E}}
\def\IF{\relax{\rm I\kern-.18em F}}
\def\IH{\relax{\rm I\kern-.18em H}}
\def\II{\relax{\rm I\kern-.17em I}}
\def\IN{\relax{\rm I\kern-.18em N}}
\def\IP{\relax{\rm I\kern-.18em P}}
\def\IQ{\relax\,\hbox{$\inbar\kern-.3em{\rm Q}$}}
\def\bfzero{\relax\,\hbox{$\inbar\kern-.3em{\rm 0}$}}
\def\IR{\relax{\rm I\kern-.18em R}}
\def\ZZ{\relax\ifmmode\mathchoice
{\hbox{\cmss Z\kern-.4em Z}}{\hbox{\cmss Z\kern-.4em Z}}
{\lower.9pt\hbox{\cmsss Z\kern-.4em Z}}
{\lower1.2pt\hbox{\cmsss Z\kern-.4em Z}}\else{\cmss Z\kern-.4em
Z}\fi}
\def\IU{\relax\,\hbox{$\inbar\kern-.3em{\rm U}$}}
\def\bfone{\relax{\rm 1\kern-.35em 1}}

\def\rmd{{\rm d}}

\def\Coe#1.#2.{{#1\over #2}}

\def\coe#1.#2.{\relax{\textstyle {#1 \over #2}}\displaystyle}

\def\to{\rightarrow}
\def\notin{\hbox{{$\in$}\kern-.51em\hbox{/}}}

\def\del{\partial}

\newcommand{\ba}{\begin{eqnarray}}
\newcommand{\ea}{\end{eqnarray}}

\csname @addtoreset\endcsname{equation}{section}
\begin{document}

\begin{titlepage}
\begin{flushright}
KUL-TF-99/42\\
DAMTP 1999-155\\
LPTENS.99/45\\
hep-th/9912049
\end{flushright}
\vspace{.5cm}
\begin{center}
{\Large\bf Isometric embedding of BPS branes in flat spaces with
two times}\\
\vfill
{\large Laura Andrianopoli $^a$, Martijn Derix $^a$,
 Gary W. Gibbons $^b$ \footnote{ Permanent address DAMTP, University of Cambridge},\\[2mm]
  Carlos Herdeiro $^c$,
 Alberto Santambrogio $^a$\\[2mm] and
Antoine Van Proeyen $^{a}$ \footnote{ Onderzoeksdirecteur, FWO, Belgium} }\\
\vfill

{\small\center
$^a$ Instituut voor Theoretische Fysica, Katholieke
 Universiteit Leuven,\\
Celestijnenlaan 200D B-3001 Leuven, Belgium
\\ \vspace{6pt}
$^b$ Laboratoire de Physique Th\'eorique de l'Ecole Normale Sup\'erieure, \\
24 Rue Lhomond, 75231 Paris Cedex 05, France \footnote{ Unit\'e Mixte de Recherche du
Central National de la Recherche Scientifique et de l'Ecole Normale
Sup\'erieure}
\\ \vspace{6pt}
$^c$ D.A.M.T.P., University of Cambridge,\\ Silver Street, Cambridge CB3
9EW, U.K.
}
\end{center}
\vfill
\begin{center}
{\bf Abstract}
\end{center}
{\small
We show how non-near horizon $p$-brane theories can be obtained from
two embedding constraints in a flat higher-dimensional space with two
time directions. In particular, this includes the construction of
$D3$-branes from a flat 12-dimensional action, and $M2$- and $M5$-branes from
13 dimensions. The worldvolume actions are determined by constant
forms in the higher dimension, reduced to the usual expressions
by Lagrange multipliers. The formulation affords insight into the
global aspects of the spacetime geometries and makes contact with recent work
on two-time physics.
} \vspace{2mm}
\end{titlepage}


\section{Introduction}

It is well known that the description of an anti-de~Sitter
($AdS$) space in $d$
dimensions is facilitated by embedding it in a flat $(d+1)$-dimensional
space with two time directions. The embedding is obtained by one
constraint and gives a manifestly $SO(d-1,2)$ symmetric
description of the global properties of the geometry. Any choice of
coordinates on the $d$-dimensional manifold will break this manifest symmetry.

An even more obvious example of extending the dimension for a better
understanding of the geometry is the description of the sphere
geometry $S^d$ in a $(d+1)$-dimensional flat space.

The two examples above are combined in \cite{conffads}, where the
$AdS_{p+2}\times S^{n-1}$ near-horizon geometry of $p$-branes in
$D=p+n+1$ dimensions is described starting from a flat
$(D+2)$-dimensional space. Two constraints are imposed, which,
respectively, reduce $p+3$ dimensions to the $AdS_{p+2}$ manifold and
$n$ dimensions to the $S^{n-1}$-sphere. The Born--Infeld actions for
the near-horizon theories of various branes are obtained from those
in a flat $(D+2)$-dimensional space by adding terms with two Lagrange
multipliers, imposing the embedding constraints. The Wess--Zumino
(WZ) terms in the actions are obtained from a $(p+2)$-form, which is
integrated over a $(p+2)$-dimensional manifold which has the
worldvolume as its boundary.

It is interesting therefore to ask whether the
full spacetime metric of a brane may be isometrically embedded in
${\IE}^{D,2}$. Apart from its possible relevance to the existence of
exotic theories with two times, it is possible that isometric
embeddings of $D$- and $M$-brane metrics as submanifolds of flat
space may have some technical advantages for quantizing particles or
strings moving in these backgrounds, since one may think of them as
free particles or strings subjected to nonlinear constraints. One
might try to implement Dirac's procedure for quantization with
constraints.

We will generalize the constructions of \cite{conffads} for non-near-horizon
brane geometries. We will show that even if the geometry is not a product of an
$AdS$ with a sphere, the brane geometry can be embedded in a flat
$(D+2)$-dimensional space with two time directions. The two
constraints are in this case not expressed on separate coordinates of
the embedding space, but involve a non-trivial mixing of the
coordinates. Also the forms for the WZ terms are obtained in this
picture. For that construction, we follow \cite{hew} where it is proposed
that a $p$-brane evolving in a space with two times should couple to
a $(p+3)$-form field strength. The field strength is contracted to a
$(p+2)$-form which can be used for the WZ term. To do this contraction
we will have to introduce an extra vector field which will be of an elegant form
only in the
near-horizon limit.

One may wonder whether the
whole geometry cannot be embedded with just one extra dimension and
why we need  two time directions in the embedding space.
First of all it has been shown \cite{eise} that the embedding of
a surface in a flat space
of co-dimension 1 imposes, by use of the Einstein equations of motion, that the
surface has constant curvature, if the surface has dimensionality
$d>2$.
This corresponds to the familiar cases of the embedding
of spheres and (anti-)de~Sitter manifolds in flat spaces with only one more dimension.
Therefore, in order to embed a brane background, we need at least two extra dimensions.

To determine the signature of the metric of the embedding space we use the following argument.
An interesting aspect of brane spacetimes is that they
are not globally hyperbolic\footnote{A space is called
globally hyperbolic if it
possesses a Cauchy surface.}. According to Penrose \cite{Penrose},
a global isometric
embedding into a flat space with one time, i.e.\ into ${\IE}^{n,1}$,
is not possible for a spacetime which is not globally hyperbolic. One
needs at least two times. Penrose's argument is essentially that the
restriction of the time coordinate $X^0$ of ${\IE}^{n,1}$ to the
embedded spacetime $M$ would serve as a time-function on $M$, i.e.\ a
function which increases along every future-directed timelike
curve. Moreover, if the embedding is suitably regular, the level sets
(constant time slices)
would actually serve as Cauchy surfaces on M, implying global hyperbolicity.
No such obstruction arises
for embeddings into flat spacetimes with more than one time.

What we will describe in this paper is therefore a minimal embedding of general brane backgrounds
in flat spaces with two extra dimensions and $(D,2)$ signature.
\vskip 5mm

In the case of particles in brane backgrounds, the constraints determining the
embedding in a space  with two time directions can be studied in the Hamiltonian formalism leading
to canonical quantization.
We first treat these constraints as second-class constraints,
 finding the corresponding system of Dirac brackets.
Furthermore, as discussed in \cite{bars,bars2}, particle systems in spaces with
 two times
can also be associated with first-class constraints closing an
$Sp(2)$ algebra (the local version of the $Sp(2)$ relating coordinates and momenta in phase space).
 We will obtain these constraints. However, it turns out that
the action for a particle in a brane background can be written
as a gauge-fixed form of the action constructed in \cite{bars}
 only in the case of a conformally flat metric, i.e.\ in the near-horizon limit.
\vskip 5mm

In section~\ref{ss:embedding} we give the embedding of the geometry,
first in general and then comment on the near-horizon approximation,
and on connecting regions separated by coordinate singularities.
The worldvolume actions will be constructed in
section~\ref{ss:braneaction}.
The essential step in that section is the construction of the forms.
First, general results for the electric field strengths are given,
before completing the construction for the cases $D3$, $M2$ and $M5$
separately.
In section~\ref{ss:particle} we will give the Dirac analysis of
constraints in the
particle case and make connection with
the works of \cite{bars}.


\section{Embedding: The geometry}
\label{ss:embedding}
In this section we describe the embedding of an $SO(n)$ invariant
$p$-brane in a $(D+2)=(n+p+3)$-dimensional spacetime. We will
obtain the embedding by demanding that the known metric of the brane
be obtained from a flat $(D,2)$ metric.
Thus, we demand that the embedding is isometric\footnote{An
embedding is said to be isometric at a point $p$ iff
$g(U,V)=\overline{g(f_{*}U,f_{*}V)}$, where $(M,g)$ is the manifold to
embed, ($\bar{M},\bar{g}$) the embedding space, $f$ is the embedding
($f:(M,g)\rightarrow (\bar{M},\bar{g})$), $f_{*}$ is the differential
of $f$, and $U$ and $V$ are two elements of $T_{p}M$ \cite{goenner,friedman}. This basically means
that length and angles are preserved in mapping from $T_{p}M$ to
$T_{p}\bar{M}$. When chart-induced bases are used both on $M$ and
$\bar{M}$, we have for the isometric condition
$g_{\alpha\beta}=X^{A}{}_{,\alpha}X^{B}{}_{,\alpha}\bar{g}_{AB}$.}.
With an ansatz where the
$D+2$ flat coordinates are written in terms of a particular mixing
of hyperspherical and horospherical coordinates, 3 of these coordinates
are left as unknown functions of the coordinate $r$ describing the
distance from the brane. This leads to two constraints, as we first
show for a rather general type of metric in $D$ dimensions with
$SO(p,1)\times SO(n)$ symmetry. Then we specialize to the metric for
non-dilatonic $D$- and $M$-branes. In the third subsection, we look
to the structure beyond the horizon to obtain an insight in the global
structure.

\subsection{The general case}\label{ss:generalcase}
The $p$-brane original spacetime is $D$-dimensional, and the
geometry can generally be described by a metric of the form
\begin{equation}
\rmd s^{2}=A(r)^{2}\left[ -\rmd t^{2}+\rmd x_{p}.\rmd x_{p}\right]+B(r)^{2}\,\rmd r^{2}+
C(r)^{2}\,\rmd \Omega_{n-1}^{2},
\label{metricbra}
\end{equation}
where $\rmd x_p. \rmd x_p$ is the $p$-dimensional spacelike part on the
worldvolume, and $\rmd \Omega_{n-1}^{2}$ is the $n$-sphere metric. The
metric has manifest Poincar\'e invariance on the worldvolume, as well as
 $SO(n)$ invariance in the transverse space.

Now we consider the embedding space, for which we consider
Cartesian coordinates $X^{M}$, with
 $M=0,\ldots ,D+1$, which we divide as follows:
\begin{equation}
  X^M=\left\{ \begin{array}{ll}
    X^\mu \qquad  & \mu =0,\ldots ,p \\
    X^{p+1} &   \\
    X^{p+2} &   \\
    X^\alpha  & \alpha=p+3,\ldots ,D+1.
  \end{array}\right.
\label{XMcomp}
\end{equation}
The flat metric with signature $(D,2)$ can be written as
\begin{equation}
\rmd s^{2}=-(\rmd X^{0})^{2}+(\rmd X^{1})^{2}+\cdots +(\rmd X^{p+1})^{2}-(\rmd X^{p+2})^{2}
+\cdots+ (\rmd X^{D+1})^{2}.
\end{equation}

{}From the viewpoint of the embedding, the $(D+2)$-dimensional coordinates
$X^M$ are
 the embedding functions. Hence, the $D$-dimensional geometry will be described
 by two constraints in the $(D+2)$-dimensional coordinates, in the same way that a
sphere $S^{2}$ embedded in  {$\mathbb{R}^{3}$} is described by one constraint in
 the 3-dimensional coordinates, namely the equation $X^{2}+Y^{2}+Z^{2}=R^2$
 in Cartesian coordinates. We now describe one way to obtain these two
constraints. We start by making a change of coordinates in the $(D+2)$-dimensional
 spacetime so as to make manifest a subgroup $SO(p,1)\times SO(n)\subset
SO(p+n+1,2)$. This is achieved by using a mixture of hyperspherical
 and horospherical coordinates $\{\rho ,z,x^\mu ,\beta ,n^\alpha \}$,
 as follows:
\begin{eqnarray}
&& X^{p+2}-X^{p+1}=\frac{\rho}{z},\nonumber\\
&& X^{p+2}+X^{p+1}= \rho z+\frac{\rho}{z}x^{\mu}x_{\mu},\nonumber\\
&& X^{\mu}=\rho \frac{x^{\mu}}{z}\nonumber,\\
&& X^{\alpha}=\beta n^{\alpha},
\label{embori}
\end{eqnarray}
where $n^{\alpha}$ parametrize\footnote{The $n^{\alpha}$ can be seen as the
Cartesian coordinates in $R^{n}$ so that they define the unit sphere $S^{n-1}$
 as $\sum_{\alpha}(n^{\alpha})^{2}=1$ and hence, the metric on this sphere,
 $\rmd \Omega_{n-1}^{2}$ is  given by the Euclidean metric on $R^{n}$
restricted to the hyperspherical hypersurface $\Sigma$. The relation
between $n^{\alpha}$ and the usual hyperspherical angular coordinates
$(\theta,\phi_{1},\ldots ,\phi_{n-2})$, is, for example, for $p=3$:
$n^{6}=\sin(\theta)\sin(\phi_{1})\sin(\phi_{2})\sin(\phi_{3})\sin(\phi_{4})$,
$n^{11}=\sin(\theta)\sin(\phi_{1})\sin(\phi_{2})\sin(\phi_{3})\cos(\phi_{4})$,
$n^{10}=\sin(\theta)\sin(\phi_{1})\sin(\phi_{2})\cos(\phi_{3})$,
$n^{9}=\sin(\theta)\sin(\phi_{1})\cos(\phi_{2})$,
$n^{8}=\sin(\theta)\cos(\phi_{1})$,
$n^{7}=\cos(\theta)$. } the sphere $S^{n-1}$.
With this change of coordinates, the metric reads
\begin{equation}
\rmd s^{2}=-\rmd \rho^{2}+\frac{\rho^{2}}{z^{2}}\left[\rmd x^\mu \,\rmd x_\mu +\rmd z^{2}\right]
+\rmd \beta^{2}+\beta^{2}\,\rmd n ^\alpha \,\rmd n^\alpha.
\label{metricemb}
\end{equation}
In the comparison between (\ref{metricemb}) and (\ref{metricbra}) we
identify $\rmd x^\mu \,\rmd x_\mu$ with $-\rmd t^2+\rmd x_p.\rmd x_p$ and $\rmd n^\alpha \,\rmd n^\alpha
$ with $\rmd \Omega _{n-1}^2$. Then $\beta $, $\rho $ and $z$ are
functions of $r$ to be determined. The comparison gives
\begin{eqnarray}
&& \beta=C(r) \nonumber\\
&& \frac{\rho}{z}=A(r)\nonumber\\
&&
-\rmd \rho^{2}+\frac{\rho^{2}}{z^{2}}\,\rmd z^{2}+\rmd \beta^{2}=B(r)^{2}\,\rmd r^{2}.
\label{ident}
\end{eqnarray}
These encode the two aforementioned constraints, since we reduce from
three degrees of freedom ($\rho,z,\beta$) to one ($r$). The
differential equation can further be written as
\begin{equation}
\frac{C'^{2}-B^{2}}{A'}=(\rho z)' \equiv F',
\label{df}
\end{equation}
where the prime denotes differentiation with respect to $r$.
 Hence, we can impose the constraints while making the coordinate transformation
 (or, in other words, while defining the embedding functions) by replacing (\ref{embori}) by
\begin{eqnarray}
X^- &\equiv& X^{p+2}-X^{p+1}=A(r)\nonumber\\
X^+ &\equiv& X^{p+2}+X^{p+1}= F(r)+A(r)x^{\mu}x_{\mu}\nonumber\\
X^{\mu}&=&A(r)x^{\mu}\nonumber\\
X^{\alpha}&=&C(r) n^{\alpha}.
\label{embedfun}
\end{eqnarray}

We can, furthermore, express the constraints in terms of the $X^{A}$
coordinates only. Denoting the inverse function with an overbar,
i.e.\ $\bar{f}(f)=f(\bar{f})={\rm identity}$, we can write
$r=\bar{A}(X^{p+2}-X^{p+1})$. Thus, our two constraints are
\begin{eqnarray}
\phi_1 (X^+,X^-,X^\mu) &=& X^-X^+-X^{\mu}X_{\mu}-
X^- F(\bar{A}(X^-))=0\nonumber\\
\phi_2 (X^-,X^\alpha) &=& \sum_{\alpha}(X^{\alpha})^{2}-
\left[ C(\bar{A}(X^-))\right]^{2}=0.
\label{constraints}
\end{eqnarray}
These constraints are thus determined by the functions $A$, $C$ and
$F$. The latter is determined up to a constant by (\ref{df}) in terms
of $A$, $B$ and $C$. Note that so far there is no definition of the
radial variable $r$. We can use different parametrizations;
for example, it will turn out that in some cases it is useful to take $A$ or $C$ itself
as the radial variable. Thus in the first case one may take the
equations here with $A(r)=r$, and in the second case $C(r)=r$.
In the standard brane cases, the functions $A$, $B$ and $C$ will take
the form of some harmonic function to some power
in the transverse space to the brane. We will further adopt the name
$r$ for that transverse coordinate, use just the name $A$ for the
parameter in the first mentioned parametrization and use $R$ for the
radial coordinate such that $C(R)=R$.

\subsection{Non-dilatonic $D$- and $M$-branes}

{}From now on we will assume that the functions $A$, $B$ and $C$ are
indeed harmonic functions in $n$ dimensions with a flat limit at
$r\rightarrow \infty$. For non-dilatonic $D$- and $M$-branes, they are of the
following form:
\begin{eqnarray}
H &= &  \left(1+\frac{1}{r^{\kappa}}\right)\nonumber\\
A(r) &=& H^{-1/(p+1)}
 \nonumber\\
B(r) &=& H^{ 1/\kappa}\nonumber\\
C(r)&=& r H^{ 1/\kappa}\nonumber\\
\kappa &\equiv &n-2=D-p-3.
\label{harmonic}
\end{eqnarray}
Here {\it a~priori} $r>0$ and $r=0$ corresponds to the horizon, but we will
come back to this in section~\ref{nondilaton}. The values for $p$, $\kappa$ and
\begin{equation}
  w=\frac{p+1}{\kappa}
\label{defw}
\end{equation}
for these branes are summarized in table~\ref{nondil}.
\begin{table}[th]
  \begin{center}
\label{tab1}
    \begin{tabular}{|l|lll|}
\hline
 & $p$ & $\kappa $ & $w$ \\
\hline
$D3$ & 3 & 4 & 1  \\
$M2$ & 2 & 6 & $\frac12$  \\
$M5$ & 5 & 3 & 2  \\
\hline
    \end{tabular}
     \end{center}
\caption{\it The non-dilatonic branes}
\label{nondil}
\end{table}
Under these conditions, the
second constraint in (\ref{constraints}) implies
\begin{equation}
\sum_{\alpha}(X^{\alpha})^{2}=\frac{(X^-)^{-2w}}
{\left[(X^-)^{-(p+1)}-1\right]^{2/\kappa}}.
\end{equation}

With an explicit form for the functions $A$, $B$ and $C$ we can evaluate the function $F$. Using (\ref{df}) we
obtain
\begin{equation}
F'(r)=- w r^{1-\kappa}
(1+r^{-\kappa})^{\frac{2}{\kappa} + \frac{1}{p+1} -1} (1+2r^\kappa),
\end{equation}
which can be integrated to give (up to a constant)
\begin{equation}
F(r)=- \frac{w}{\kappa} \left[ B_\frac{r^\kappa}{r^\kappa
+1}\left(\frac{-1}{p+1}, 1-\frac{2}{\kappa}\right) + 2
B_\frac{r^\kappa}{r^\kappa+1} \left( \frac{p}{p+1}, -\frac{2}{\kappa}\right)
\right]\!.
\label{fofr}
\end{equation}
Here we used the incomplete beta function
\[ B_x(a,b)=\int_0^x  t^{a-1} (1-t)^{b-1} \,\rmd t=a^{-1}x^a \,{}_2F_1(a,1-b;a+1;x), \]
which is defined for $0 < x \leq 1$. This means that $F(r)$ is well
defined in the region $r>0$, which is what we were looking for.


\paragraph{The near-horizon approximation.}
It is well known that the isometry group of $AdS_{n}$, i.e.\ $SO(n-1,2)$,
 acts as the
 conformal group on an $(n-1)$-dimensional Lorentzian manifold (in particular, on
 its conformal boundary). It has also been known for some time
 \cite{interp} that $AdS$
 spacetimes arise as the geometry of some BPS branes in the near-horizon
limit. For instance, for the $M2$-, $M5$- and $D3$-branes, the near-horizon geometry
 is $AdS_{p+2}\times S_{D-p-2}$. Therefore, if we study these branes on the
backgrounds of their own near-horizon geometries, conformal field theories on
 the branes worldvolumes should arise. In  \cite{conffads}, such
 a study was performed by embedding the near-horizon supergravity solutions in
 a $(D+2)$-dimensional spacetime. As we discussed in the previous section, for
 the three mentioned solutions the
 geometry takes a form like (\ref{metricbra}) with $A$, $B$ and $C$ given by
(\ref{harmonic}).
For small $r$, we obtain
\begin{equation}
  B=\frac{C}{r}\sim \frac{1}{r},\qquad A\sim
 r^{1/w},\qquad F\sim w^2 r^{-1/w}\,
\label{ABCnh}
\end{equation}
so that for the three cases, the embedding
 functions (\ref{embedfun}) reduce to those used in \cite{conffads}.


\subsection{Passing through the horizon}

Using the embedding (\ref{embedfun})
 we can now study the global properties of the
brane geometries. Before considering the higher-dimensional $D$- and
$M$-branes, let us first look at the easier and lower-dimensional
example of the extreme Reissner--Nordstr{\o}m (RN) black hole (a large
list of embedding functions for other solutions of general relativity
is given in \cite{rosen}).

\subsubsection{Example: extreme Reissner--Nordstr{\o}m black hole.}

The RN black hole fits our general embedding scheme with $D=4$ and
$p=0$, $\kappa =1$, $w=1$. Here, rather than working with the radial
variable $r$ as in (\ref{harmonic}), we use the variable $R$, which,
as mentioned at the end of section~\ref{ss:generalcase}, has the
property $C(R)=R$. Then the functions $A$ and $B$ are given by
$A(R)=B(R)^{-1}=1-1/R$. The variable $R$ is just shifted with
respect to $r$, as $R\equiv r+1$, and the horizon is now at
$R=1$, and $R=0$ corresponds to the singularity.
Using (\ref{df}), we then find
\[ F_{RN}(R)=\frac{1}{R-1} -3R - R^2 - 4 \log |R-1| .\]

\begin{figure}[H]
\begin{center}
\leavevmode
\epsfxsize=6.5cm
\epsfbox{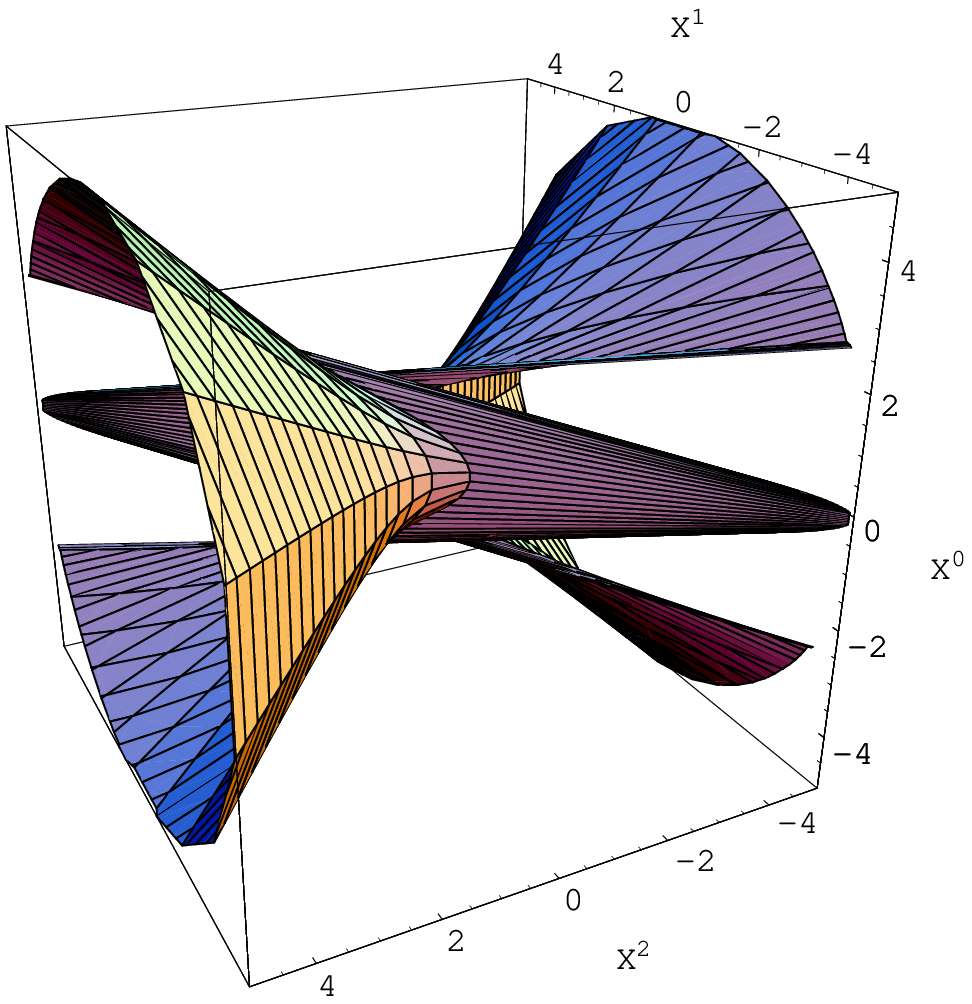}
\epsfxsize=6.5cm
\epsfbox{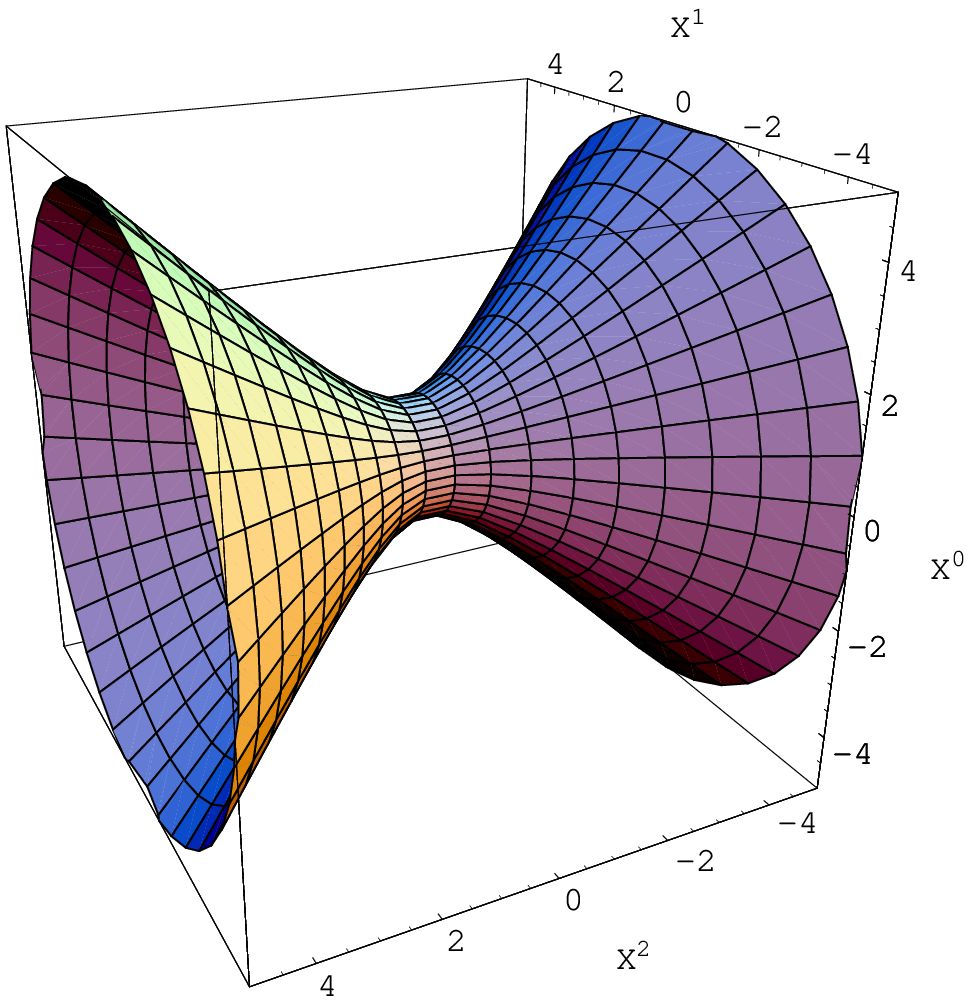}
\caption{\it $AdS_2$. Left, horospherical coordinates ($X^-=A$,
$X^+=1/A-A{}t^2$, $X^0=A{}t$) and right, hyperspherical
coordinates ($X^0=\sqrt{1+u^2}\cos{v}$, $X^1=\sqrt{1+u^2}\sin{v}$,
$X^2=u$). Note $X^\pm=X^1 \pm X^2$.}\label{ads2}
\end{center}
\end{figure}

The near-horizon limit of the extreme RN metric is $AdS_2 \times
S^2$. $AdS$ spaces are naturally defined by their embedding.
Figure~\ref{ads2} shows $AdS_2$ using two different parametrizations.
Note that the horospherical embedding seems to have problems at
$X^-=0$, because this parametrization goes bad as $A=X^- \to 0$. This
is just an artefact of the parametrization and is not a feature of the
embedded surface itself, as can be seen from the other
parametrization. The holes in the left figure are due to limits in
the range of $t$ and $r$. For example, we have to start from a finite $r>0$
in order to have no singularities in the functions.

\begin{figure}[H]
\leavevmode
\centerline{
\epsfysize=8.5cm
\epsfbox{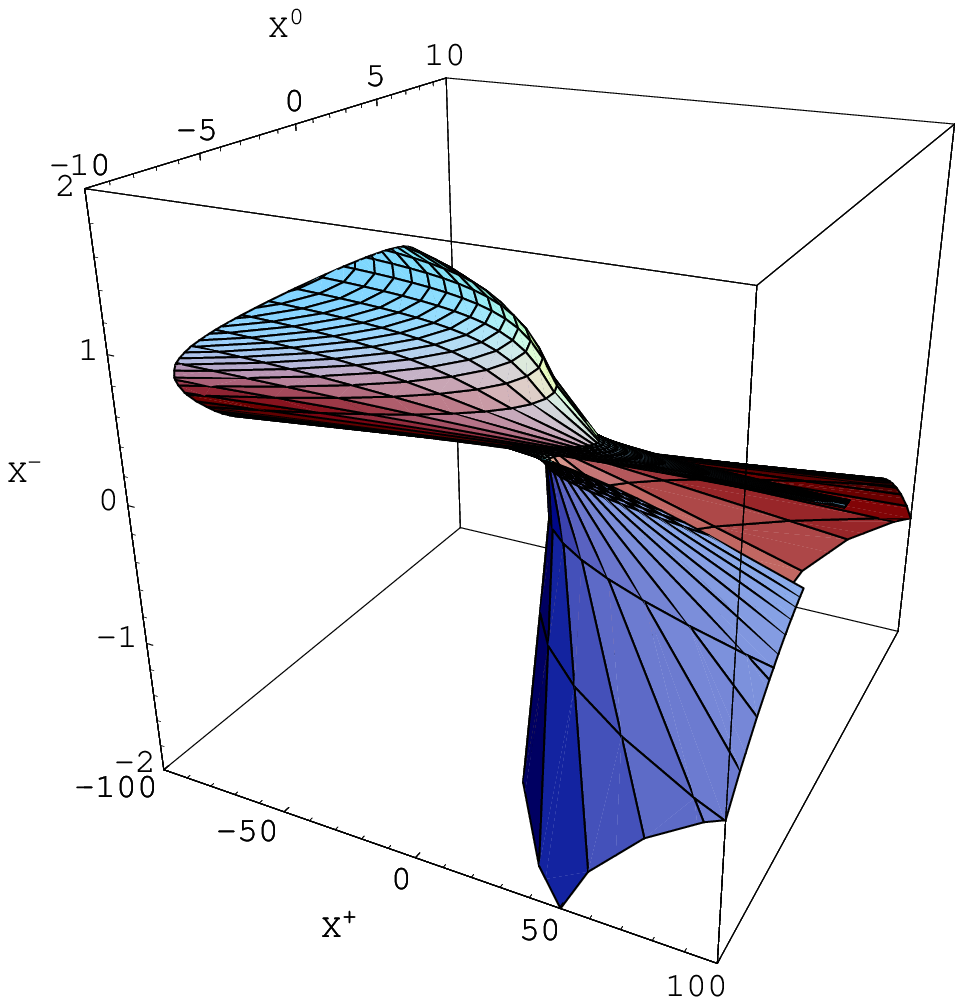}
}\\
\centerline{
\epsfysize=8.5cm
\epsfbox{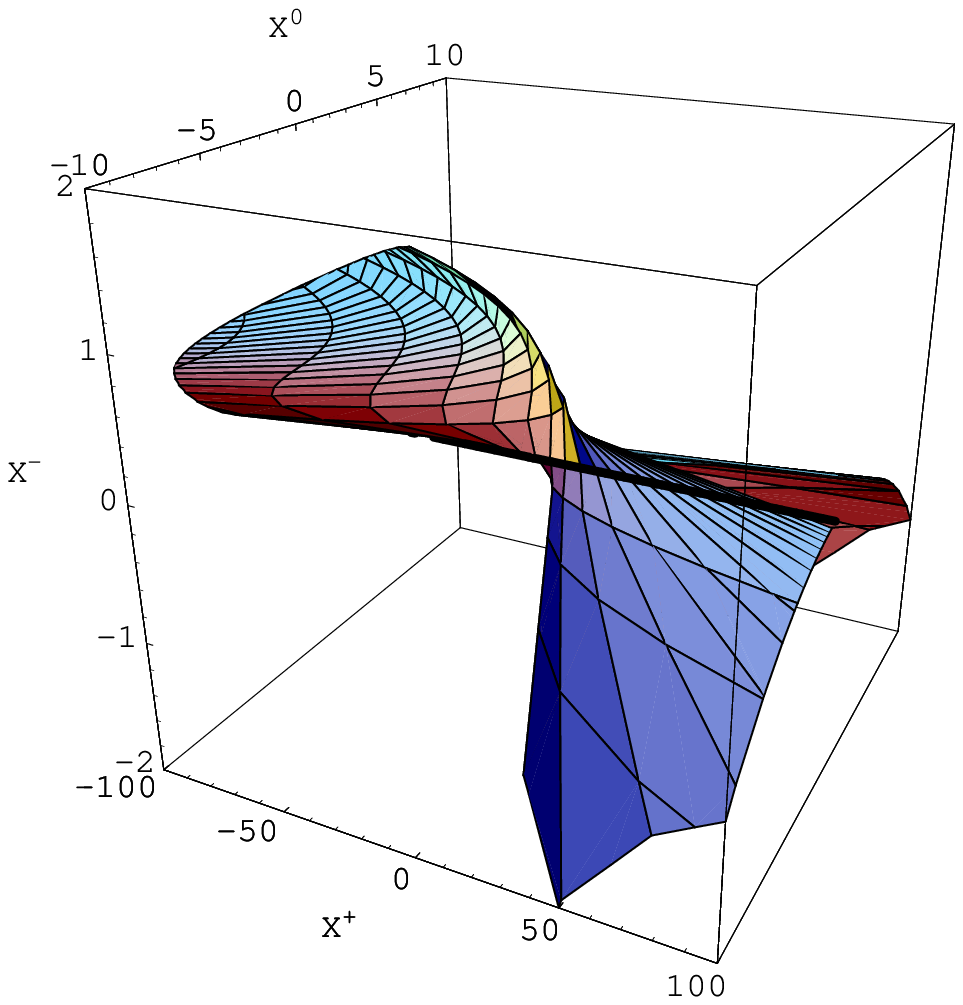}
}
\caption{\it Extreme Reissner--Nordstr{\o}m black hole. Top, parametrized
by $R$ and $t$. Bottom, in advanced Finkelstein coordinates $R$ and
$v$ (with $v=t-R^*$, where $R^*=\int \frac{B(R)}{A(R)}\,\rmd R$). The fat line indicates the
horizon. Recognize the two different regions: on the left the
asymptotically flat region I ($X^->0$) and on the right the interior
region II ($X^-<0$) connected in the AdS throat. Compare the constant
$v$ (bottom) with constant $t$ lines (top) which correspond to
infalling lightlike geodesics. In the $v$ parametrization they pass
through the horizon into the interior region and end at the singularity
instead of diverging at the horizon.}
\label{rn}
\end{figure}

The entire Reissner--Nordstr{\o}m black hole geometry can be drawn in a similar
fashion using parametrization (\ref{embedfun}) as is shown in
figure~\ref{rn}. Note that we used a slightly different orientation
as in the $AdS_2$ pictures.

We can read off the following global features from the picture. The geometry
consists of 2 distinct regions: region I is the asymptotically flat region for
$R>1$ which corresponds to $X^->0$. For big $R$ the surface flattens
and $X^- \to 1$, which is the flat limit. Region
II is the region inside the horizon ($X^- <0$). The singularity ($R=0$)
corresponds to $X^- \to -\infty$.
The two regions are  connected in an {\it $AdS$ throat} (cf
figure~\ref{ads2}). It seems
that these two
regions are disconnected, the constant time lines all diverge near
$X^-=0$ and never cross the horizon. As in the $AdS$ case, this is just
an artefact of the parametrization. Actually, we know that the
near-horizon geometry is equivalent to $AdS_2$, which has no problems
at its `horizon'.

\begin{figure}[H]
\leavevmode
\centerline{
\epsfysize=7cm
\epsfbox{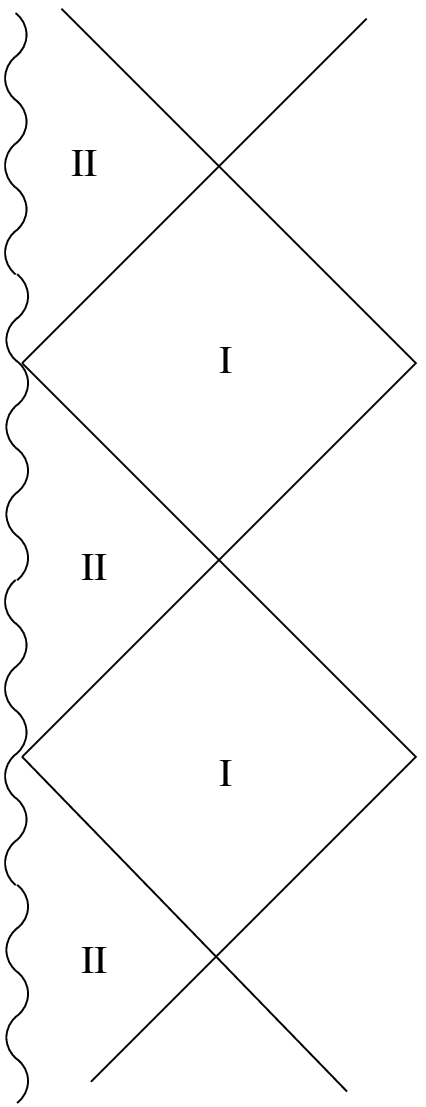}
\hskip 1cm
\epsfysize=7cm
\epsfbox{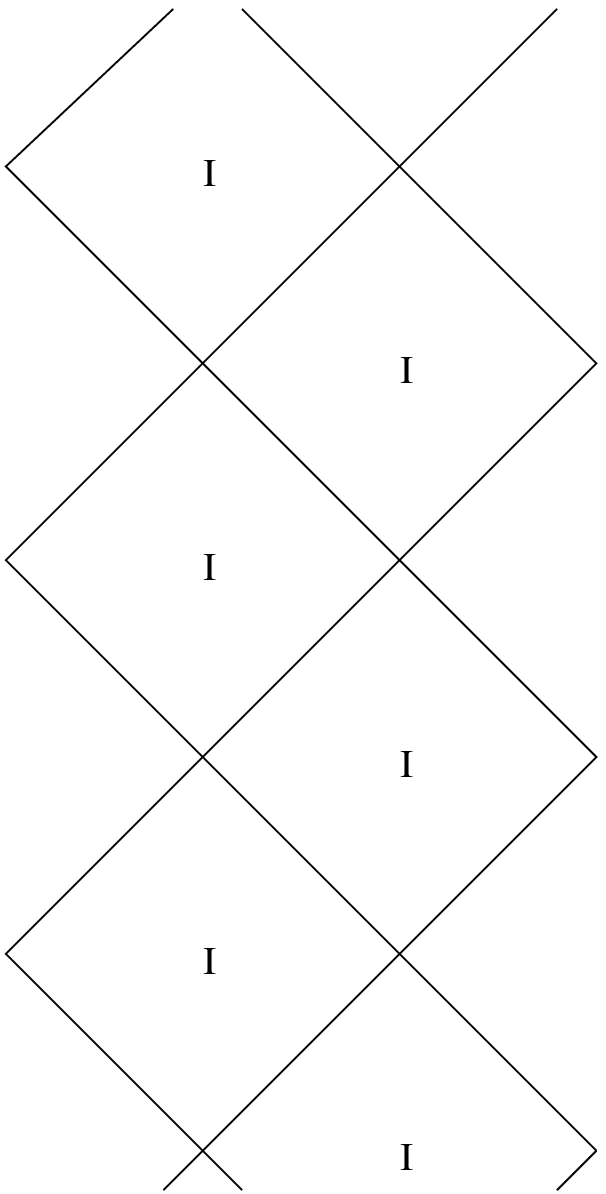}
}
\caption{\it Penrose--Carter diagrams. Left, $p$ even (including the
extreme Reissner--Nordstr{\o}m black hole); right, $p$ odd. The different regions I and II
are indicated.
\label{diagrams}}
\end{figure}

One of the features of $AdS$ spaces is that they admit closed timelike
curves, which can be clearly seen in figure~\ref{ads2}. The usual
remedy for this is to consider the covering space $CAdS$ instead
of $AdS$ itself. Looking at figure~\ref{rn} we see that our RN
black hole geometry suffers from the same problem, it admits closed timelike
curves (remember that $X^0$ and $X^2$ both are timelike
directions). Again this is remedied by considering the covering
space. The result of this of course is that the space then consists of
multiple universes. It is this covering space that is depicted in the
familiar Penrose--Carter diagram for the extreme RN black hole
(figure~\ref{diagrams}).

In textbooks one usually shows that the horizon of a black hole is
regular by adopting special coordinates in which you can follow
the geodesics through the horizon inside the black hole, the so-called
advanced (or retarded) Finkelstein coordinates. We can do the
same thing with our embedding as is shown in figure~\ref{rn}.

\subsubsection{Non-dilatonic branes.}\label{nondilaton}

As discussed in \cite{GHT}, the general brane solution case
(\ref{harmonic}) can be divided in two classes: $p$ odd or $p$ even.

\begin{figure}[H]
\leavevmode
\centerline{
\epsfysize=9cm
\epsfbox{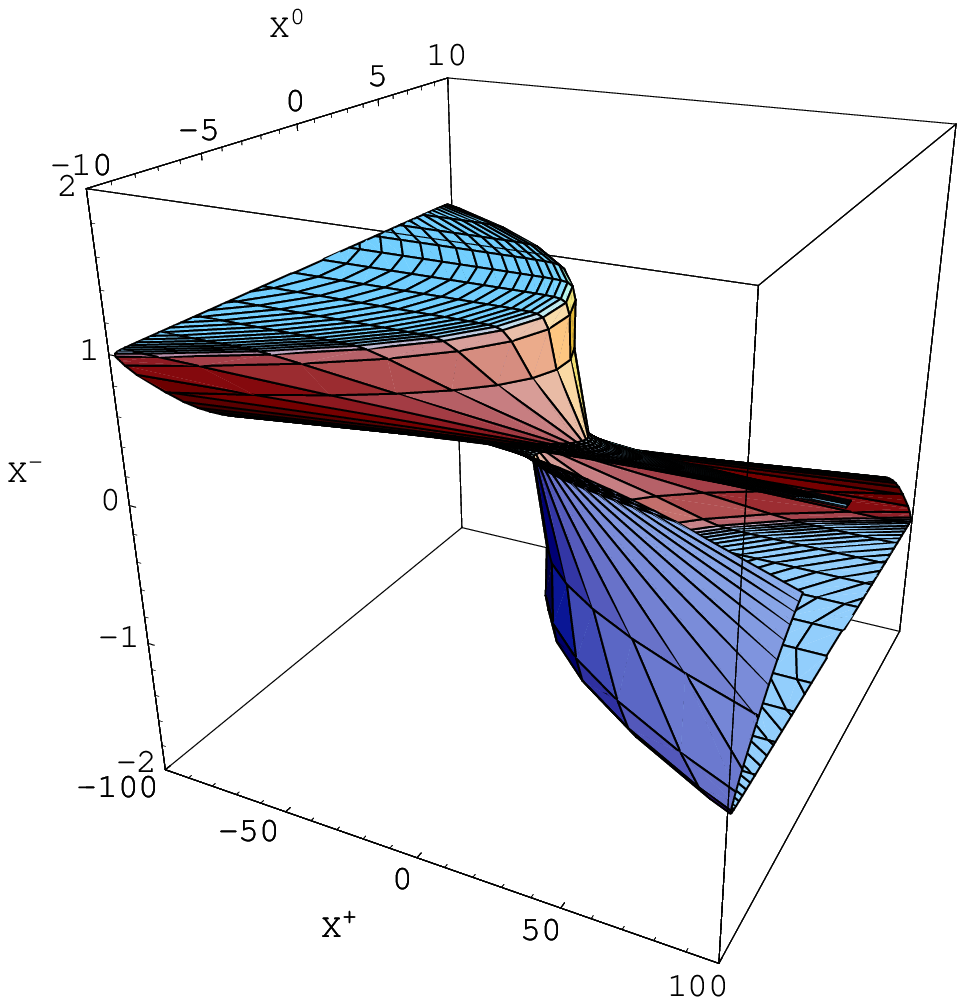}}
\centerline{
\epsfysize=9cm
\epsfbox{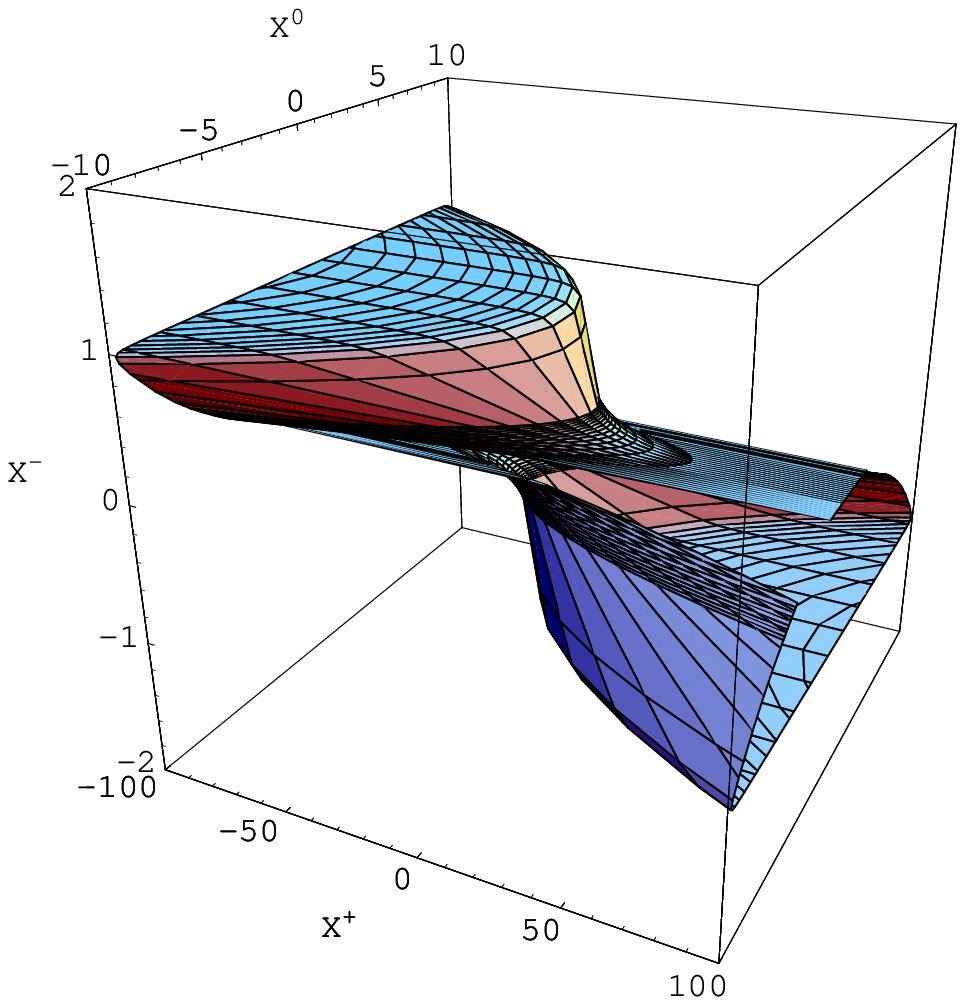}
}
\caption{\it $p$ odd branes. a: (top) $D3$ and b: (bottom)
$M5$. Two asymptotically flat regions connected in $AdS$.}
\label{d3m5}
\end{figure}

Let us first consider the $p$ \emph{odd} case. In the region $r>0$, the exterior
region, the function $A(r)$ is analytic and positive and vanishes as
$r\to 0$. If we take $A$ to be our new radial variable instead of $r$,
we see that $A$ can be continued through the horizon to negative $A$
\cite{GHT}. The range of $A$ is from $-1$ to~1.
The analytic extension of the metric is
\begin{equation}
  \rmd s^2= A^2\,\rmd x_\mu \,\rmd x^\mu +\left(1- A^{p+1}\right) ^{-2/\kappa}
  \left[ w^2\left(1- A^{p+1}\right) ^{-2}A^{-2}\,\rmd A^2+\rmd\Omega ^2\right] ,
\label{dsA}
\end{equation}
which is even in $A$. This leads to
\begin{equation}
F(A) =-  \frac{w}{\kappa}  (\mbox{sign }A) \left[ B_{A^{p+1}}
\left(\frac{-1}{p+1}, 1-\frac{2}{\kappa}\right) + 2
B_{A^{p+1}} \left( \frac{p}{p+1}, -\frac{2}{\kappa}\right)
\right]\!.
\end{equation}
The embedding functions (\ref{embedfun}) are then odd in $A$.
This means that
the embedded space is symmetric around the horizon and completely
non-singular. (Up to a factor, $A$ corresponds to $\omega$ in
\cite{GHT}, where the same continuation was discussed.)
The Penrose diagram for these $p$ odd brane geometries is shown in
figure~\ref{diagrams}.
For the non-dilatonic branes, $D3$ and $M5$ fit this picture. The
embedding (figure~\ref{d3m5}) shows these features nicely.
It is clearly visible that there is no interior region, just two symmetric
`exterior' regions connected in the AdS throat as was expected from
the Penrose diagram.

In the $p$ \emph{even} case, the metric and embedding functions are neither
even nor odd. It is useful in this case to adopt so-called
Schwarzschild coordinates, which are defined by $R^\kappa=r^\kappa
+1$. In these coordinates the horizon (which is still a coordinate
singularity) is at $R=1$. At $R=0$ there is a true curvature
singularity (the radial variable used in the Reissner--Nordstr{\o}m
example actually was a Schwarzschild variable).
Expressed in this coordinate, $A(R)$ can be continued
through the horizon into negative $A$ and its range is
$\{-\infty,1\}$. As already stated in \cite{GHT}, the Penrose diagram
for these spaces is equivalent to the extreme Reissner--Nordstr{\o}m diagram
(see figure~\ref{diagrams}).

\begin{figure}[H]
\leavevmode
\centerline{
\epsfxsize=7.8cm
\epsfbox{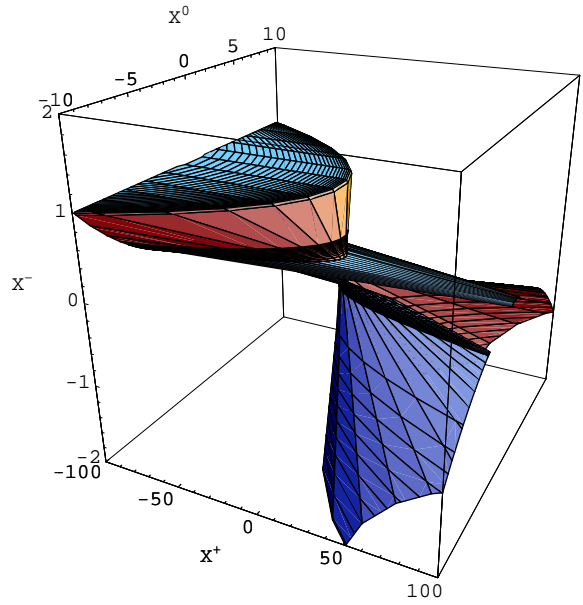}
}
\caption{\it $p$ even: $M2$. It has the same structure as the RN
black hole}
\label{figm2}
\end{figure}

The embedding of the $M2$-brane metric illustrates these features
(figure~\ref{figm2}). The expression (\ref{fofr}) of $F$ is  well
defined only in the region $R>1$. It is not possible to find a
continuous
expression for $F$ valid in both regions ($0<R\leq 1$ and
$R>1$). Nevertheless, a continuous embedding is obtained using in the interior region
\[ F(R)=\frac{w}{\kappa} \left[ B_{R^\kappa}\left(
\frac{1}{p+1} + \frac{2}{\kappa}, -\frac{1}{p+1}\right) -2
B_{R^\kappa}\left( \frac{1}{p+1} +
\frac{3}{\kappa},-\frac{1}{p+1}\right) \right]\!. \]
Note that the global structure of the $M2$ indeed resembles that of
the extreme RN black hole (cf figures~\ref{figm2} and \ref{rn}).

As in the RN case, we can directly read off some of the global
properties from the figures. Again the spaces admit closed
timelike curves or, as it is put in \cite{GaryWrap}, they are wrapped
in time. The alternative is of course again taking the covering space,
resulting in an infinite stack of
connected universes.

\begin{figure}[H]
\leavevmode
\centerline{
\epsfysize=8cm
\epsfbox{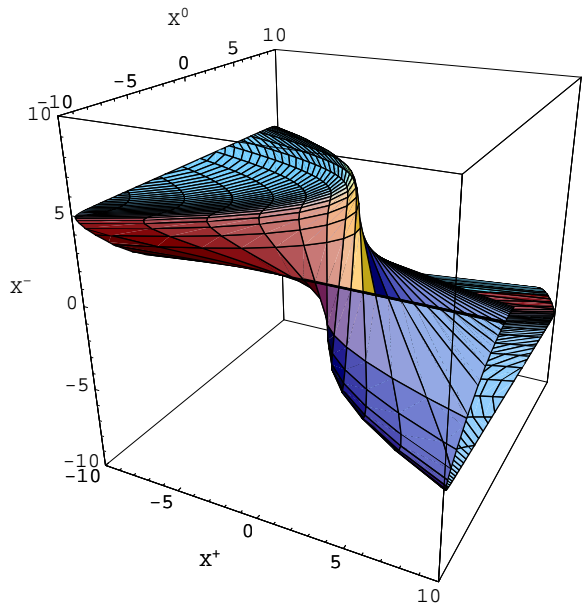}
}
\caption{\it $M5$-brane metric  parametrized by $r$ and $v=t+R^*$.
Constant $v$ lines (infalling
lightlike geodesics) pass the horizon into the next region.
}
\label{m5adv}
\end{figure}

To show that the horizon is completely regular, we can reparametrize
the embedding in Finkelstein coordinates as in the case of the RN
black hole. Figure~\ref{m5adv} shows part of the embedding of the
$M5$-brane metric using these
Finkelstein coordinates. The constant $v$ lines ($v$ is the advanced
Finkelstein time coordinate $v=t+R^*$ as defined in most textbooks)
correspond to infalling lightlike geodesics and they clearly pass through the
horizon into the next region.


\section{The brane action}\label{ss:braneaction}
We would like to write the action of a brane placed in the background of
other branes using the embedding in the  $D+2=n+p+3$ spacetime.
A typical (schematic) form of the action is
\begin{eqnarray}
S_{p+1}&=& \int_W \rmd ^{p+1}\xi\, \sqrt{-\det {\cal G}_{\mu \nu }} +
\int_{B\, (W=\partial B)} \Omega _{(p+2)}  \nonumber\\
&+& \int_W \rmd^{p+1}\xi\, [\lambda_1 \phi_1(X^+,X^-,X^\mu) +
\lambda_2 \phi_2(X^-,X^\alpha)],
\label{action}
\end{eqnarray}
where $W$ is the $(p+1)$-dimensional worldvolume of the brane.
The expression for ${\cal G}_{\mu \nu }$ differs for  each case.
For example, for $Dp$-branes ${\cal G}_{\mu\nu} \equiv  \partial _\mu X^M \partial _\nu
 X^N \eta_{MN}+{\cal F}_{\mu\nu}$, with
 ${\cal F}_{\mu\nu}$ the field strength of the gauge field living on the worldvolume
of the brane. The fields  $\lambda_1,
\lambda_2$ are two Lagrange multipliers implementing the constraints
(\ref{constraints}). $\Omega _{(p+2)}(X^M)$ is a function of
the forms coupling to the brane, such that it reduces to the
appropriate Wess--Zumino term when projected on the physical
hypersurface determined by the constraints.
The explicit form of  $\Omega_{(p+2)}$ will be determined, for the
non-dilatonic $D3$-, $M2$- and $M5$-brane cases, in the next subsections.

\subsection{Embedding the field strength}
Let us now study the embedding in the $(D+2)$-dimensional space of the
field strengths appearing in the Wess--Zumino term of the action (\ref{action}).
We will follow the suggestion of \cite{hew}.
The idea would be that a brane (extended in $p$ spatial directions) fluctuating
in a spacetime with two times
should evolve in both the time directions, and therefore couple to
a ($p+3$)-form field strength.
We assume therefore that the $(D+2)$-dimensional theory can be coupled to
rank $p+3$ electric field strengths $K_e$, and to rank $n$ magnetic
field strengths $K_m$, coupling to branes with $p$ spacelike
 directions evolving in 2 times.
We then ask that the rank $p+2$ field strengths $F$
coupled to $p$-branes in $D$-dimensional supergravity  be  a restriction of
 $K_e$ ($K_e +K_m$ for the $D3$-brane case) to the
$D$-dimensional hypersurface $\Sigma$.
This ansatz is the most natural one for the $D3$-brane,
because in this case the $10$-dimensional self-dual field strength is extended
to a self-dual field strength in $12$ dimensions.
If there were a supergravity theory in
$D=12$, the bosonic configuration with flat
$(10,2)$-space and a constant self-dual  field strength
would solve the equations of motion. This is obvious for
the Maxwell equation (there can be no Chern--Simons terms built from a
5-form potential in 12 dimensions and so the Maxwell equation would take
the standard form), but for the Einstein equations it is only true because
the field strength is self-dual. In a $D$-dimensional spacetime with
Euclidean or Kleinian signature (i.e.\ zero or two times) a self-dual field
strength has a vanishing energy--momentum tensor for $D=4$ mod 4.
(For Lorentzian signature it is $D=2$ mod 4.)
What this would mean is that the ten-dimensional $D3$-brane solution would
just be the projection to a complicated hypersurface of an almost trivial
12-dimensional supergravity solution.

Let us start by analysing how an electric $(p+2)$-form
 field strength $F^{(p+2)}$ coupling to a $p$-brane becomes
embedded in the $(D+2)$-dimensional space.
Our aim is to obtain $F$ as a restriction of a $(p+3)$-form $K^{(p+3)}$ to the
$D$-dimensional hypersurface $\Sigma$.

A general non-dilatonic brane is described in $D$ dimensions by the fields
 \cite{d3brane}
\begin{eqnarray}
\rmd s^{2}&=&H^{-2/(p+1)}\left[ -\rmd t^{2}+\rmd x_1^{2}+ \cdots +\rmd x_p^{2} \right]
 + H^{2/\kappa}\left[ \rmd r^{2} +r^2 \,\rmd
 \Omega_{D-p-2}^{2}\right]\!,
\nonumber\\
G_{01\ldots p}&=&-H^{-1}=-A^{p+1},
\nonumber\\
\Phi&=&\mbox{constant}=0,
\label{fields}
\end{eqnarray}
with the notation of section~\ref{ss:embedding}.
The only non-vanishing components of the field strength $F=\rmd G$, are
those related to the components given above by antisymmetry.
Taking that into account, we can write the (electric) field strength as
\begin{equation}
F=-(p+1) A^p A' \,\rmd r\wedge \rmd t\wedge \rmd x_1\wedge \cdots
\wedge \rmd x_p,
\label{fieldstrength}
\end{equation}
where a prime denotes differentiation with respect to $r$.

\vskip 5mm

To find the embedding,
we start by considering a constant  $(p+3)$-form in $D+2$ dimensions
\begin{equation}
K_e=\frac{p+1}{(p+3)!}\epsilon_{\mu'_{0}\ldots \mu'_{p+2}}
\,\rmd X^{\mu'_{0}}\wedge \rmd X^{\mu'_{1}}\wedge \cdots \wedge \rmd X^{\mu'_{p+2}}.
\label{p+3form}
\end{equation}
Primed indices run over $\mu'=0,\ldots,p+2$. In order to obtain a rank $(p+2)$ field strength, we contract $K_e$ with a vector
 field $V$, with components $V=V^M(\frac{\partial}{\partial X^M}$),
 which we take to be arbitrary (there is a sign ambiguity in this contraction;
 we chose to make it on the left, i.e.\ $A_{\mu'_{1}\ldots \mu'_{p+2}}=
V^{\mu'_{0}}K_{\mu'_{0}\ldots \mu'_{p+2}}$). Such a contraction yields
\begin{equation}
K_e(V)=\frac{p+1}{(p+2)!}\epsilon_{\mu'_{0}\ldots \mu'_{p+2}}V^{\mu'_{0}}
\,\rmd X^{\mu'_{1}}\wedge \cdots \wedge \rmd X^{\mu'_{p+2}}.
\label{p+2form}
\end{equation}
Then, we reduce the resulting $(p+2)$-form to the $D$-dimensional hypersurface
 by using the embedding functions (\ref{embedfun}),
\begin{eqnarray}
K_e(V)|_{\Sigma} &=&
 \half(p+1) A'A^{p+1}  \,\rmd r\wedge \rmd t \wedge \rmd x_1 \wedge
\cdots \wedge \rmd x_p
 \nonumber\\ &&\times
 \times \left[ 2V^{\mu}x_{\mu}+ V^{+}\left(\frac{F'}{A'}-x^{\mu}x_{\mu}\right)
-V^{-}\right]\!,
\label{hev}
\end{eqnarray}
where we have defined $V^{\pm} \equiv V^{p+2}\pm V^{p+1}$.
Next we impose that $K_e(V)|_\Sigma=F$.
To determine $V^{M}$,  we use the ansatz
 $V^M=(V^{\mu'},V^\alpha )=(\alpha(r)X^{\mu'},V^\alpha )$. Because
$K_e$ only has components in the longitudinal directions, $V^\alpha$
stays undetermined. When the field strength also includes a magnetic
part, this $V^\alpha$ comes into play, as we will see in the next section.
 It follows that, in order for (\ref{hev}) to match with (\ref{fieldstrength}),
 $\alpha (r)$ has to obey
\begin{equation}
\alpha(r)\left(\frac{AF'}{A'}-F\right)=-\frac{2}{A}.
\label{alpha}
\end{equation}
We use (\ref{df}) to determine $\alpha(r)$
\begin{equation}
\alpha(r)=\frac{2}{AF+w^2 C^2 (2C^{\kappa}-1)}.
\end{equation}
The general form of the vector field in terms of the $(D+2)$-dimensional
coordinates is
\begin{equation}
V^{\mu'}=\frac{2X^{\mu'}}{w^2 (X^{\alpha}X_{\alpha})
\left[2(X^{\alpha}X_{\alpha})^{\kappa/2}-1 + w^{-2}
\right] -X^{M}X_{M}}.
\end{equation}
If we take the near-horizon limit, with use of (\ref{ABCnh}) we
find $\alpha(r)\sim {1}/{w^2}$, which matches with the results in
\cite{conffads} (see formula (2.17) in that paper).

\subsection{$D3$-brane embedding}
The 10-dimensional Wess--Zumino term is the integral of the self-dual field strength $F$
that couples to the $D3$-branes solution of the type~IIB supergravity theory.
For the 12-dimensional theory we construct a self-dual 6-form $K$, i.e.\
\begin{equation}
\star{K}\wedge K=\eta_{12}|K|^{2},
\end{equation}
where $\eta_{12}$ is the volume form on the 12-dimensional spacetime.
Our aim is to obtain $F$ as a restriction of $K$ to the 10-dimensional surface $\Sigma$.
The $D3$-brane is described by the fields (\ref{fields}) introduced in the previous
subsection, with $p+1=\kappa=4$, $D=10$.
In this case, the only non-vanishing components of the field strength
 $F=\rmd G$ are those related to the components in (\ref{fields}) by either
 antisymmetry or
 self-duality. Taking that into account we can write the (anti)self-dual field strength as
\begin{equation}
F=H' H^{-2}\,\rmd t\wedge \rmd x\wedge \rmd y \wedge \rmd z \wedge
\rmd r
+ H'H^{-1/2}\sqrt{|g|} \,\rmd \theta\wedge \rmd \phi_{1}\wedge
\rmd \phi_{2}\wedge
\rmd \phi_{3} \wedge \rmd \phi_{4},
\end{equation}
where a prime denotes differentiation with respect to $r$.
In terms of the embedding functions, the last expression reads
\begin{equation}
F =- 4 A'A^{3}\,\rmd t\wedge \rmd x\wedge \rmd y \wedge
\rmd z \wedge\rmd r + 4 \omega_{(5)},
\label{fieldstr}
\end{equation}
where $\omega_{(5)} \equiv
\sin(\theta)^{4}\sin(\phi_{1})^{3}\sin(\phi_{2})^{2}\sin(\phi_{3})
 \,\rmd \theta
\wedge \rmd\phi_1 \wedge \cdots \wedge \rmd\phi_4$  is the volume form on the
unit 5-sphere.
\vskip 5mm

To find the embedding,
we start by considering a constant self-dual six-form in 12 dimensions
\begin{equation}
K=\frac{4}{6!}(\epsilon_{\mu'_{0}\ldots \mu'_{5}}\,\rmd X^{\mu'_{0}}\wedge \rmd X^{\mu'_{1}}
\wedge \cdots \wedge \rmd X^{\mu'_{5}} + \epsilon_{\alpha_{1}\ldots \alpha_{6}}\,\rmd X^{\alpha_{1}} \wedge \rmd X^{\alpha_{1}}
\wedge \cdots \wedge \rmd X^{\alpha_{6}}).
\label{sixform}
\end{equation}
We have written $K$ in a manifestly $SO(4,2)\times SO(6)$ invariant form.
In order to obtain a rank-5 field strength, we contract, as we have done for the
general electric case, $K$ with a vector field $V$,
 with components $V=V^M(\frac{\partial}{\partial X^M}$).
Such a contraction yields
\begin{equation}
K(V)=\frac{4}{5!}(\epsilon_{\mu'_{0}\ldots \mu'_{5}}V^{\mu'_{0}}\, \rmd X^{\mu'_{1}}\wedge \cdots \wedge \rmd X^{\mu'_{5}} + \epsilon_{\alpha_{1}
\ldots \alpha_{6}}
V^{\alpha_{1}} \,\rmd X^{\alpha_{2}}\wedge \cdots \wedge \rmd X^{\alpha_{6}}).
\label{fiveform}
\end{equation}
Then, we reduce the resulting 5-form to the 10-dimensional hypersurface by using the embedding functions
(\ref{embedfun}). By requiring the matching $K(V)|_{\Sigma}=F$, we obtain the constraints for our
vector field $V$. The resulting 5-form $K(V)$ is the looked-for Wess--Zumino
form $\Omega_5$.

Let us analyse separately the two terms in the right-hand side of
(\ref{fieldstr})--(\ref{fiveform}).
The electric part has already been studied in the general case in the
previous subsection. In this case it gives $V^{\mu'}=\alpha(r) X^{\mu'}$
with
\begin{equation}
  \alpha (r)=\frac{2}{2 (X^{\alpha}X_{\alpha})^3  -X^{M}X_{M}}.
\end{equation}

The angular part in (\ref{sixform}) can be rewritten in terms of the radial
coordinate $r$ and of the angular coordinates $\theta$,$\phi_i$ ($i=1,\ldots
,4$) as
\begin{eqnarray}
\frac{1}{6!} \epsilon_{\alpha_{1}\ldots \alpha_{6}}\,\rmd X^{\alpha_{1}} \wedge \cdots
\wedge \rmd X^{\alpha_{6}} &=&
C' C^5 \sqrt{g_{S_5}} \,\rmd r \wedge \rmd\theta\wedge
\rmd\phi_{1}\wedge
\cdots
\wedge \rmd\phi_{4} \nonumber\\
&=& C' C^5 \,\rmd r \wedge \omega_5,
\end{eqnarray}
where $g_{S_5}$ is the determinant of the metric of the unit 5-sphere in polar coordinates..
In order for the second term in (\ref{fiveform}) to match with the second
 term in (\ref{fieldstr}), we have to require that the vector $V$ points in the
 radial direction when decomposed on the $r,\theta , \phi_i$ basis, that is
\begin{equation}
V^\alpha \frac{\del }{\del X^\alpha} \equiv V^\alpha \frac{\del r }{\del X^\alpha}
\frac{\del }{\del r}.
\label{radialv}
\end{equation}
This gives
\begin{equation}
\frac{1}{5!} \epsilon_{\alpha_{1}\ldots \alpha_{6}}V^{\alpha_{1}}  \,\rmd X^{\alpha_{2}}
\wedge \cdots
\wedge \rmd X^{\alpha_{6}}=
C' C^5 V^\alpha \frac{\del r }{\del X^\alpha} \omega_5.
\end{equation}
The matching with (\ref{fieldstr}) requires
\begin{equation}
V^\alpha \frac{\del r }{\del X^\alpha}=(C' C^5)^{-1},
\end{equation}
which is solved, with the ansatz $V^{\alpha}=\epsilon(r) X^{\alpha}$,
by\footnote{
We have used the relation $C^2(r)=X^\alpha X_\alpha$, from which
$\frac{\del r }{\del X^\alpha}=\frac{\del r }{\del C^2(r)}\frac{\del C^2 }{\del X^\alpha}=(CC')^{-1}X^\alpha$.}
\begin{equation}
V^\alpha=C^{-6} X^\alpha=(1+r^4)^{- 3/2}X^\alpha.
\end{equation}
\vskip 5mm

We notice that $\epsilon(r=0)=\alpha(r=0)=1$, so that in the near-horizon
 approximation we have $V^{M}=X^{M}$ and $K(V)$ becomes the potential for
 the 12-dimensional self-dual form $K$ (up to a constant),
 making contact with the work in \cite{conffads}.
The general form of the vector field in terms of the 12-dimensional coordinates is
\begin{equation}
V^{\mu'}=\frac{2X^{\mu'}}{2(X^{\alpha}X_{\alpha})^{3}-X^{M}X_{M}}, \quad\quad
V^{\alpha}=\frac{X^{\alpha}}{(X^{\beta}X_{\beta})^{3}}.
\end{equation}

\subsection{$M2$-brane embedding}
For $M2$ the Wess--Zumino form is the rank 4 (electric) field strength $F$.
We require that this field strength be a restriction of $K_e$
to the 11-dimensional hypersurface.

The fundamental $M2$-brane is described in 11 dimensions by the fields
(\ref{fields}) with $p+1=3$, $\kappa =6$,  and  $F=\rmd G$ an electric
 field strength, i.e.\
\begin{equation}
F=-3A' A^2\,\rmd x_0\wedge \rmd x_1\wedge \rmd x_2 \wedge \rmd r.
\label{m2}
\end{equation}

Repeating the general analysis, we require that
$F=K(V)|_\Sigma$, where $K$ is a constant 5-form in 13 dimensions
\begin{equation}
K=\frac{3}{5!}\epsilon_{\mu'_{0}\ldots \mu'_{4}}\,\rmd X^{\mu'_{0}}\wedge \rmd X^{\mu'_{1}}
\wedge \cdots \wedge
\rmd X^{\mu'_{4}}.
\end{equation}
The  vector field $V$,
 with components $V=V^M(\frac{\partial}{\partial X^M})$, turns out to be
$V^M=(V^{\mu'}, V^\alpha)=(\alpha(r)X^{\mu'}, V^\alpha)$
where
\begin{equation}
\alpha(r)=\frac{8A}{4A^{2}F+r^2(1+2r^{6})},
\end{equation}
and $V^\alpha$ stays undetermined.
Written in terms of 13-dimensional coordinates, the general form of the
vector field $V^M$ is
\begin{equation}
V^M=\left(V^{\mu'},V^\alpha \right)
=\left(\frac{8X^{\mu'}}{X^\alpha X_\alpha \left[3+2(X^\alpha X_\alpha )^3\right]
-4X^M X_M}, V^\alpha \right)\!.
\end{equation}
The $\Omega _4$ of (\ref{action}) is $K(V)$.

\subsection{$M5$-brane embedding}

Let us now perform a similar construction for the $M5$-brane background.
We start with the 11-dimensional metric describing the
 geometry of the $M5$-brane
\begin{equation}
\rmd s^2=H^{-1/3}(-\rmd t^2 +\rmd x_1^2+\cdots+ \rmd x_5^2)
+ H^{2/3} ( \rmd r^2 + r^2 \,\rmd \Omega_{(4)} ),
\end{equation}
with $H=1 + 1/r^3$.
The $M5$ is a solitonic solution of 11-dimensional supergravity
\cite{dsg}, and is coupled in a magnetic way to the
 4-form
\begin{equation}
F^{(4)} =\rmd A^{(3)}=H' H^{-1/2}\sqrt{|g|}\,
\rmd\theta \wedge \rmd\phi_1 \wedge \cdots \wedge \rmd\phi_3=3 \omega _{(4)}
\end{equation}
and in an `electric' way to the 7-form
\begin{equation}
F^{(7)}=\star F^{(4)} + A^{(3)}\wedge F^{(4)}=
 -H^{-2} H' \, \rmd x^0 \wedge \cdots \wedge \rmd x^5 \wedge \rmd r.
\label{m5f}
\end{equation}

The expression (\ref{m5f}) for the electric coupling to $M5$ has the same form
(\ref{fieldstrength})
as the general electric field strength discussed previously, so we can apply
the same procedure.
Analogously to the $M2$ case, we start from a constant 8-form
$K^{(8)}$ in $d=13$
and contract it with an arbitrary vector (let us call it $V_1^M$)
 to a 7-form
\begin{equation}
K^{(8)}(V_1)=\frac{6}{7!} \epsilon_{\mu_0' \ldots \mu_7'} V_1^{\mu_0'}
\,\rmd X^{\mu_1'} \wedge \cdots \wedge \rmd X^{\mu_7'} .
\end{equation}
The requirement that $K(V_1)|_\Sigma=F$ fixes the vector $V_1^M=(V_1^{\mu'},
V_1^\alpha )$ to be of the form
$V_1^M=(\alpha (r) X^{\mu'}, V_1^\alpha )$ with
\begin{equation}
\alpha(r)=\frac{2}{ - X^M X_M + X^\alpha X_\alpha [ 8 ( X^\beta
X_\beta )^{3/2} - 3 ]} .
\end{equation}

Extra complications arise when we want to describe the $M5$ propagating
in the background of the antisymmetric gauge fields $F^{(4)}$ and
$F^{(7)}$ of 11-dimensional supergravity.
Indeed, the 3-form field strength ${\mathcal H}$
\cite{m5}  living on the worldvolume
of the $M5$ couples to the above fields giving the following expression
for the  Wess--Zumino term:
\begin{equation}
\left.\Omega_7 \right|_\Sigma=
 F^{(7)} -\half {\cal H}\wedge F^{(4)}.
\label{7form}
\end{equation}
It is therefore this last expression that has to be embedded in
the 13-dimensional space.
We will follow and generalize the discussion in
 \cite{conffads}, where an analogous
treatment of the $M5$ Wess--Zumino term has been given for the near-horizon case.

As for the near-horizon limit,
we only need the embedding of the fields $F^{(7)}$
and $F^{(4)}$, ${\mathcal H}$ being a field living on the physical
worldvolume of the brane, only defined over $\Sigma$.

To find the embedding of $F^{(4)}$ we can proceed in a way completely
similar to the magnetic part of the $D3$-brane field strength:
we start from a constant 5-form in 13 dimensions,
\begin{equation}
 K^{(5)}=\star K^{(8)}=\frac{6}{5!} \epsilon_{\alpha_{1}\ldots \alpha_{5}}\,\rmd X^{\alpha_{1}} \wedge \rmd X^{\alpha_{1}}
 \wedge \cdots \wedge \rmd X^{\alpha_{5}}.
\label{5form}
\end{equation}
In order to obtain a rank-4 field strength we contract, as we have done for the
other cases, $K$ with a vector field $V_2$,
 with components $V_2=V_2^{M}(\frac{\partial}{\partial X^{M}}$).
Such a contraction yields
\begin{equation}
K^{(5)}(V_2)=\frac{6}{4!} \epsilon_{\alpha_{1}\ldots \alpha_{5}}
V_2^{\alpha_{1}} \,\rmd X^{\alpha_{2}}\wedge \cdots \wedge \rmd
X^{\alpha_{5}}.
\label{4form}
\end{equation}
Then we reduce the resulting 4-form to the 11-dimensional hypersurface by using the embedding functions (\ref{embedfun})
\begin{equation}
K^{(5)}(V_2)|_{\Sigma}=C' C^4 V_2^\alpha
\frac{\partial r}{\partial X^\alpha} \, \omega_4.
\end{equation}
By requiring $K^{(5)}(V_2)|_{\Sigma}=F^{(4)}$,
we can solve for the
vector field $V_2$, which turns out to be (following the same arguments as for $D3$)
$V_2^M=(V_2^{\mu'}, V_2^\alpha )$, with $V_2^{\mu'}$ free and
\begin{equation}
V_2^\alpha=C^{-5}X^\alpha.
\end{equation}
Note that the two vectors $V_1^M$ and $V_2^M$ have constrained components on
orthogonal subspaces ($V_1$ on the indices $\mu'$ and $V_2$ on the
indices $\alpha$), so we can use for the projection the single vector
$V^M=(V_1^{\mu '},V_2^\alpha )$.

We then find, analogously to \cite{conffads},
that the 7-form appearing in the embedded Wess--Zumino
term has the following form:
\begin{equation}
\Omega_7=K^{(8)}(V)-\half {\mathcal H}
\wedge K^{(5)}(V),
\end{equation}
which, restricted to the hypersurface $\Sigma$, reduces to the closed form
(\ref{7form}).


\section{Particles in a brane background}
\label{ss:particle}
Now we would like to analyse  the behaviour of a probe particle put in a
$p$-brane background.
In particular, we would like to see whether the above construction (an embedding in a
 higher-dimensional flat space) can shed new light on the dynamics of the
system.

We can interpret the particle in a background as a constrained
Hamiltonian system and apply to it the standard rules of constrained
quantization.
In particular, the constraints (\ref{constraints}) selecting the physical
 hypersurface
can be seen to  appear in that context as second-class primary constraints.

Recently, however, an approach  has been developed by Bars and collaborators \cite{bars}
 for studying Hamiltonian systems
embedded in higher-dimensional spaces with two times. Here the constraints
 follow from an $Sp(2)$ gauge
 symmetry internal to the particle system\footnote{It is a local version
 of the $Sp(2)$ global group relating coordinates and conjugate momenta
 in phase space.}. They are therefore
first-class constraints (instead of second class).
In that way it was shown that any physical system that can be written as a gauge-fixed form of
a particular $(D+2)$-dimensional action with $Sp(2)$ gauge symmetry (the action
(2) in
 \cite{bars} (BDA action)) has, in fact, a (maybe hidden) $SO(D,2)$ off-shell
 symmetry.

Having found an embedding of brane backgrounds in a flat spacetime with two
time directions,
it is natural to wonder whether a system given by a particle in a brane
background could be written as a gauge-fixed form of the BDA action (and has therefore,
if this were the case, a hidden $SO(D,2)$ invariance).

It turns out that, in general, this is not true:
 the action for a particle in a brane background can be written as
gauge-fixed form of the BDA action only when the background metric is conformally flat,
that is only in the asymptotic regions of the background.
The details of our application of the construction in \cite{bars} are given in section~\ref{ss:1clconstr}. Before doing this, let us describe how the resolution of
the constraints works when using the second-class constraints approach.
%

\subsection{Hypersurface as a set of second-class constraints}
 We start from a free Hamiltonian in the embedding space,
\begin{equation}
H_0=P^M P_M=-4 P_+ P_- + P_\mu^2 + P_\alpha^2
\end{equation}
on which we impose the following two constraints:
\begin{eqnarray}
\phi_1 &=& - X^+ X^- + (X^\mu)^2 + X^- F(X^-) , \\
\phi_2 &=& (X^\alpha)^2 - C(X^-)^2,
\end{eqnarray}
to constrict the motion of the particle to the hypersurface
(note that $\eta_{+-}=-\frac{1}{2}$, $\eta^{+-}=-2$).
Note that $F(X^-)$ should be interpreted as $F( \bar  A(X^-) )$. Following the general procedure of quantizing constrained Hamiltonian systems, we add these primary constraints to the
Hamiltonian,
\begin{equation}
H_*=H_0 + u^i \phi_i .
\end{equation}
In order to find out whether there are any secondary constraints, we calculate the bracket of the primary constraints with the
Hamiltonian,
\begin{equation}
\dot \phi_i=\{ \phi_i , H_* \} \approx \{ \phi_i, H_0 \} + u^j \{\phi_i , \phi_j \} ;
\end{equation}
the brackets are defined by $\{X^M , P_N \}=\delta^M{}_N$. From this we find the following two secondary constraints:
\begin{eqnarray}
\chi_1 &=& X^{\mu '} P_{\mu '} - (X^- F)' P_+ , \nonumber\\
\chi_2 &=& X^\alpha P_\alpha + (C^2)' P_+,
\end{eqnarray}
where a prime denotes differentiation with respect to $X^-$.
Taking the bracket of the new constraints with the Hamiltonian $H_*$ we find
two equations that allow us to fix the coefficients $u^i$:
there are no tertiary constraints.

These four constraints are all what is called second class, because there is
no subset of them that has zero bracket with all other constraints.
In order to deal with them we make use of the Dirac analysis
\cite{dirac}. Let us introduce
the so-called Dirac brackets
\begin{equation}
\{f,g \}_D=\{ f,g \} - \{f, \varphi_a \} C^{ab} \{\varphi_b, g\} ,
\end{equation}
where $\varphi_a=(\phi_i , \chi_i)$, $a=1,\ldots, 4$ and
$C^{ab}$ is the inverse of the matrix
\begin{equation}
\Delta_{ab}=\{ \varphi_a, \varphi_b \},
\end{equation}
which is non-singular because all constraints are second class.

With these new brackets we can calculate the equations of motion
via Hamilton's  equations
\begin{equation}
\dot X^M=\{ X^M, H_0 \}_D, \qquad \dot P_M=\{ P_M , H_0 \}_D .
\end{equation}
Applying this to our case we find some very non-trivial differential equations for the $X^M$, describing the motion of a particle confined to the embedded
spacetime,
\begin{eqnarray}
&&\half \ddot X_M=\dot P_M
=\frac{1}{2C^2(X^-)^2[(C')^2-F']}\times
\Big\{[2\delta_M^{\mu '}X_{\mu '}+\delta_M^- (X^-F)']\times \nonumber\\
&&\phantom{.}\hspace{5mm}
\left\{2C^2\left[P^{\nu '}P_{\nu '}+\half (P^-)^2(X^-F)''\right]
+X^-(C^2)' \left[P^{\beta}P_{\beta}-\half (P^-)^2(C^2)''\right]\right\}
\nonumber\\
&&+[2\delta_M^{\alpha}X_{\alpha}-\delta_M^- (C^2)'] \times
\left\{X^-(C^2)'\left[P^{\nu '}P_{\nu '}+\half (P^-)^2
(X^-F)''\right] \right.
 \nonumber\\
&&\phantom{.}\hspace{45mm}\left.\left.+
2(X^-)^2F'
 \left[P^{\beta}P_{\beta}-\half
 (P^-)^2(C^2)''\right]\right\}\right\}.
\end{eqnarray}

In the near-horizon limit we have $C^2\sim 1$ and $X^-F\sim w^2$. The
equations of motion then take the much simpler form
\begin{equation}
\half \ddot X_M=\dot P_M=
\frac{2}{w^2} \delta_M^{\mu '} X_{\mu '}
(P^{\nu '}P_{\nu '}) - 2\delta_M^{\alpha} X_{\alpha}
(P^{\beta}P_{\beta}).
\end{equation}

\subsection{First-class constraint approach}\label{ss:1clconstr}

We refer, for the context discussed here, to the recent series of papers by Bars and collaborators
\cite{bars,bars2}.
There, many systems are discussed where the apparent symmetry of the action
can be enlarged to a bigger, nonlinearly realized, $SO(D,2)$
hidden symmetry with explicit time-dependent
 generators. They show that all these (very different) actions can be
written as one and the same action in $D+2$ dimensions, with $(D,2)$ signature,
where the $SO(D,2)$ is now  realized linearly.
In that context, the different actions can be seen as following from
different gauge choices
of a hidden $Sp(2)$ gauge symmetry present in the $(D+2)$-dimensional action.
In some sense, the point of view in \cite{bars,bars2}
is that every time that there is a hidden $SO(D,2)$ off-shell symmetry in
 an action,
this should indicate that there is an underlying two-times physical spacetime
controlling the system. The action considered, is just one of a
 collection
of actions all related by some sort of duality, and all derivable by the same
underlying higher-dimensional theory through a particular gauge choice.

A nice point is that, in particular, the constraints that define the embedding of the
 $AdS_5 \times S_5$
in 12 dimensions  can be seen as originating from
particular choices of the $Sp(2)$ gauge on the 12-dimensional action.
This suggests that the same could be extended to our case,
just with more general constraints.
We were, therefore, led to investigate whether that scenario also fits in
the non-near-horizon case.
\vskip 5mm

Let us consider the BDA action \cite{bars}
\begin{equation}
S_0=\int_0^\tau \rmd\tau\, (D_\tau X_i^M) \epsilon^{ij} X_j^N
\eta_{MN},
\end{equation}
where $i,j=1,2 \in Sp(2,R)$,
\begin{eqnarray}
&& X_1^M =X^M  ,\qquad  X_2^M =P^M =\frac{\partial S_0}{\partial \dot X^N}\eta ^{MN} \nonumber\\
&& D_\tau X_i=\frac{\rmd X_i}{\rmd \tau}-\frac{1}{2} A_i{}^j X_j\nonumber\\
&&A_i{}^j(\tau)\in \mbox{ Adj } Sp(2,R).
\end{eqnarray}
It can be rewritten as
\begin{equation}
S_0=\int_0^\tau \rmd\tau \,(\partial_\tau X^M  P^N - \half
 X_i^M\epsilon^{ij}A_j{}^kX_k^N)\eta_{MN}.
\label{s0}
\end{equation}
We want to see whether, with an appropriate choice of $Sp(2,R)$ gauge,
 it reduces to
 the $D$-dimensional action describing the motion of a particle in a background
generated by a brane
\begin{equation}
S=\int_0^\tau \rmd \tau
\,e^{-1}G_{mn}\frac{\rmd y^m}{\rmd \tau}\frac{\rmd y^n}{\rmd \tau},
\label{s}
\end{equation}
where
\begin{eqnarray}
y^m &=&  (x^\mu,r,n^\alpha),
\qquad n^\alpha n_\alpha =1,
\nonumber\\
\rmd s^2 &\equiv & G_{mn} \,\rmd y^m\,\rmd y^n=A^2(r) \,\rmd x^\mu \,\rmd
x_\mu + B^2(r) \,\rmd r^2 + C^2(r) \,\rmd \Omega^2_{n-1},
\end{eqnarray}
and $e^{-1}$ is the worldline einbein.
\vskip 3mm

To this end, we make use of the embedding (\ref{embedfun}), that is
\begin{equation}
X^M=\left(A(r),F(r)+A(r)x^\mu x_\mu, A(r)x^\mu , C(r)n^\alpha
\right)\!.
\end{equation}
By requiring the action (\ref{s0}) to be invariant for variations with respect to
the functions $A_i{}^j$, we obtain the three constraints:
$X^MX_M=0$,  $X^MP_M=0$, $P^MP_M=0$.

The constraint $X^MX_M=0$ gives
\begin{equation}
 A(r)F(r) =C^2(r).
\label{constr}
\end{equation}
{}From (\ref{df}) and (\ref{constr}), we can eliminate $F(r)$ and obtain
\begin{equation}
B^2 =C'^2-F'A'=\left(C'-\frac{C}{A}A' \right)^2.
\label{b2}
\end{equation}
Note this new condition relating the 3 functions $A,B,C$ appearing in the metric.

Now impose  the constraint $X^MP_M =0$, with
\begin{equation}
P_M=\left( 0, P_+, \frac{p_\mu}{A(r)}, p_\alpha \right)\!,
\end{equation}
which gives
\begin{equation}
P_+=\frac{2}{A(r)} \left( p_\mu x^\mu + C(r) n^\alpha p_\alpha
\right)\!.
\end{equation}

Let us now substitute the gauge-fixed expressions for $X^M$ and $P^M$  into
the action (\ref{s0}). We obtain
\begin{equation}
S_0=\int_0^\tau \rmd\tau \left[ \frac{\rmd x^\mu}{\rmd \tau} p_\mu +
\left(\dot C-\frac{C}{A}\dot A \right)n^\alpha p_\alpha
+ C \frac{\rmd n^\alpha}{\rmd \tau} p_\alpha - \frac{1}{2} A_{2 2}\left(
\frac{p^\mu p_\mu}{A^2} + p_\alpha p^\alpha \right)  \right]\!,
\end{equation}
where, for all functions $f(r)$, $\dot f=f' \frac{dr}{d\tau}$.
Then, after eliminating $p_\mu$, $p_\alpha$ through the equations of motion,
we obtain
\begin{equation}
S_0 \rightarrow
\frac{1}{2} \int_0^\tau \rmd \tau
\,\frac{1}{A_{2 2}}\left[A^2(r)\frac{\rmd x_\mu}{\rmd\tau}\frac{\rmd x^\mu}{\rmd \tau}+ B^2(r) \dot r^2
+ C^2(r)\frac{\rmd n_\alpha}{\rmd \tau}\frac{\rmd n^\alpha}{\rmd \tau}\right]\!,
\end{equation}
where $B$ is given by (\ref{b2}). This action has the form
(\ref{s}).

In this way we have seen
that, under condition (\ref{b2}), the action for a particle in a brane
 background has a hidden $SO(D,2)$ symmetry.
We have then found that the BDA action (\ref{s0}) reduces, by appropriate gauge fixings, to the action (\ref{s}) not for the general brane
background metric
(\ref{metricbra}) but
only when the constraint (\ref{b2})
is satisfied.

This condition is fulfilled for an $AdS \times S$ metric with equal radii for the two factors,
but not for a general brane
background, with $A,B,C$ given by the usual harmonic functions. In
particular, this is true for the near-horizon limit of $D3$, but not
for $M2$ and $M5$.
In fact, condition (\ref{b2}) follows precisely by imposing the metric
(\ref{metricbra}) to be conformally flat.
Starting from (\ref{metricbra})
and using variable $C/A$,
the metric takes the form
\begin{equation}
\rmd s^{2}=A(r)^{2}\,\rmd x^\mu \,\rmd x_\mu +\frac{B(r)^{2}}
{ \left(C'-\frac{C}{A}A' \right)^2}A(r)^{2}
\,\rmd \left( \frac{C}{A}\right) ^{2}+A(r)^{2}\left(
\frac{C}{A}\right)^{2}\,\rmd \Omega_{n-1}^{2},
\label{metricbra''}
\end{equation}
which is conformally flat when (\ref{b2}) is satisfied.

\section{Discussion}
Our aim in this paper has been to develop a global description
of the spacetime geometries of $M$- and $D$-branes
by isometrically embedding them in flat spacetimes with two extra dimensions
and two times, thus extending the ideas of \cite{conffads}.
We have gained a rather clear
global picture of the geometry, giving an insight into the structure around
coordinate singularities and in the symmetries. In particular,  the differences between $p$-branes
with $p$ even and $p$ odd,  previously pointed out in \cite{GHT},
are clearly apparent.
Like the familiar embedding of anti-de~Sitter spacetime as a quadric,
our embeddings are periodic in time.  This is consistent with some suggestions in
\cite{GaryWrap}, but one may of course always pass to a covering spacetime.
But see \cite{SW} for  an isometric embedding of the universal covering spacetime
of $AdS_2$  in three and four-dimensional Minkowski
spacetimes.

In the context of supergravity and string theory, $p$-branes are
coupled to $(p+2)$-form field strengths. An embedding of
the brane thus has to include, besides the embedding of the geometry,
a prescription for the forms in the higher-dimensional space. This is
obtained by defining constant $(p+3)$-forms in $D+2$ dimensions, and
contracting them using a vector $V$. The form of $V$ is determined by
matching the projection on the surface with the known forms for $D3$,
$M2$ and $M5$. Unfortunately, the geometric significance of the vector field $V$,
remains unclear.
In the case of an $M2$-brane it is not even
unique, since the $V^\alpha $ components are arbitrary.
A co-dimension-2 surface has a 2-dimensional normal
plane.  In the $D3$ and $M5$ cases, the vector $V$ does not lie in this 2-plane, except
in the near-horizon limit. Specifically, the normal 2-plane
is spanned by $\partial _\mu \phi _1$ and $\partial _\mu \phi _2$.
One may check that $V$ is not a linear combination of
$\partial _\mu \phi _1$ and $\partial _\mu \phi _2$.
The bosonic action for probe branes in
the embedded background (\ref{action}) is completely determined after
the construction of $V$.

One motivation for our work was the possibility of using this approach to  quantize
strings moving in these backgrounds. This we have not done, but we have indicated how Dirac's
theory of constrained systems could, in principle, be used to quantize a point particle.
One point of interest is that this would automatically build in the
periodic temporal boundary conditions of \cite{GaryWrap} showing at least that the suggestion
is mathematically consistent.
Our work also makes contact with recent ideas on physics with two times.
In particular, we see difficulties in extending the methods  of \cite{bars}
from the vicinity of the throat to the entire spacetime.

Finally, it is possible that the methods developed in this paper may
be applicable to scenarios in which one regards the universe as a
brane embedded in a higher-dimensional spacetime.

\medskip
\section*{Acknowledgments}
\noindent
We would like to thank  Renata Kallosh with whom GWG and AVP  had stimulating discussions
at the early stages of this work. We would also like to thank Paul Townsend and Walter Troost for discussions. This work is
supported by the European Commission TMR programme
ERBFMRX-CT96-0045. CH is supported by FCT (Portugal) through grant
no~PRAXIS XXI/BD/13384/97.

\end{document}